# Reversible and irreversible processes during cyclic voltammetry of an electrodeposited manganese oxide as catalyst for the oxygen evolution reaction

Javier Villalobos[1], Ronny Golnak[2], Lifei Xi[1], Götz Schuck[3] and Marcel Risch[1,4,*]

[1] Nachwuchsgruppe Gestaltung des Sauerstoffentwicklungsmechanismus, Helmholtz-Zentrum Berlin für Materialien und Energie GmbH, Hahn-Meitner Platz 1, 14109 Berlin, Germany

[2] Department of Highly Sensitive X-ray Spectroscopy, Helmholtz-Zentrum Berlin für Materialien und Energie, Albert-Einstein-Straße 15, 12489 Berlin, Germany

[3] Department of Structure and Dynamics of Energy Materials, Helmholtz-Zentrum Berlin für Materialien und Energie, Hahn-Meitner Platz 1, 14109 Berlin, Germany

[4] Institut für Materialphysik, Georg-August-Universität Göttingen, Friedrich-Hund-Platz 1, 37077 Göttingen, Germany

* marcel.risch@helmholtz-berlin.de

**Abstract**

Manganese oxides have received much attention over the years among the wide range of electrocatalysts for the oxygen evolution reaction (OER) due to their low toxicity, high abundance and rich redox chemistry. While many previous studies focused on the activity of these materials, a better understanding of the material transformations relating to activation or degradation is highly desirable, both from a scientific perspective and for applications. We electrodeposited Na-containing $MnO_x$ without long-range order from an alkaline solution to investigate these aspects by cyclic voltammetry (CV), scanning electron microscopy (SEM) and X-ray absorption spectroscopy (XAS) at the Mn-K and Mn-L edges. The pristine film was assigned to a layered edge-sharing $Mn^{3+/4+}$ oxide with Mn-O bond lengths of mainly 1.87 Å and some at 2.30 Å as well as Mn-Mn bond lengths of 2.87 Å based on fits to the extended X-ray fine structure (EXAFS). The decrease of the currents at voltages before the onset of the OER followed power laws with three different exponents depending on the number of cycles and the Tafel slope decreases from 186±48 to 114±18 mV dec$^{-1}$ after 100 cycles, which we interpret in the context of surface coverage with unreacted intermediates. Post-mortem microscopy and bulk spectroscopy at the Mn-K edge showed no change of the microstructure, bulk local structure or bulk Mn valence. Yet, the surface region of $MnO_x$ oxidized toward $Mn^{4+}$, which explains the reduction of the currents in agreement with literature. Surprisingly, we find that $MnO_x$ reactivates after 30 minutes at open-circuit (OC), where the currents and also the Tafel slope increase. Reactivation processes during OC are crucial because OC is unavoidable when coupling the electrocatalysts to intermittent power sources such as solar energy for sustainable energy production.

**Introduction**

Among the wide range of transition metal oxides used for the oxygen evolution reaction (OER), manganese oxides have received much attention over the years because their low toxicity and their high abundance (10$^{th}$ in Earth crust). Furthermore, photosystem II has an active site consisting of CaMn$_4$O$_5$, which makes manganese oxides scientifically interesting as biomimetic catalysts [1–6]. In the past years, the attention was mainly focused on simple manganese oxides [3–13], such as Mn$_3$O$_4$, Mn$_2$O$_3$, MnO$_2$ and variants with non-binding, redox-inactive cations, e.g. δ-AMnO$_2$ (A is group I/II cation) [14–16]. The activation and degradation processes of Mn oxides are less understood than the activity of these catalysts. Usually, degradation is associated with changes of the surface and bulk oxide composition, long range order (i.e. crystallinity) and/or microstructure, but it can be also caused by detachment, particle agglomeration, or blocking by oxygen bubbles [17–21]. These material modifications due to activation and degradation can be electrochemically observed as changes in overpotential at fixed current or changes in current at fixed overpotential [17,22–24]. Cyclic voltammetry (CV) is the other common electrochemical test of these processes and frequently used as an activation procedure [25–28]. This activation procedure is also known as pretreatment or conditioning, and it is usually carried out as several cycles in a limited range of voltage. The promotion of the catalytic properties after an activation process has been studied for oxides such as Ir- [26,28], Co- [27,29] and also Mn-based oxides [25,30]. Common sweep speeds for these studies are in the range of 10 to 200 mV/s, while they are slower for studies with the focus on activity. Typically, the activation is performed for 20 to 50 cycles, but in some cases, the final activated material does not form until the 200$^{th}$ cycle [31].

Understanding the processes of activation and degradation of oxides in alkaline media requires measurements complementary to electrochemical methods. Frydendal et al. [32] argue that the stability degradation of electrodeposited Mn oxides cannot be determined based on electrochemical experiments alone, which is supported by gravimetric experiments using an electrochemical quartz crystal microbalance (EQCM) and inductively coupled plasma mass spectrometry (ICP-MS) . Geiger et al. [33] recently proposed the S-number as a metric of catalyst degradation which requires the determination of the material loss, e.g. by ICP-MS measurements. Köhler et al. [1] and Baumung et al. [34] studied the activation of LiMn$_2$O$_4$ particles during CV using a rotating ring-disk electrode (RRDE) where the ring was set to a voltage sensitive to dissolved Mn.

Valuable insight into the mechanism of catalysis, activation and degradation of manganese oxides has been obtained in the last decade from spectroscopy including X-ray absorption spectroscopy (XAS) [35–43], UV-Vis spectroscopy [44–46] and Raman spectroscopy [46]. We focus on alkaline electrolytes herein. The distinction between acidic, neutral and alkaline conditions is important because Mn$^{3+}$ ions disproportionate to Mn$^{2+}$ and Mn$^{4+}$ in acidic and neutral electrolytes, while Mn$^{2+}$ and Mn$^{4+}$ ions comproportionate to Mn$^{3+}$ in alkaline electrolytes [44,45]. Nakamura and coworkers [44,45] discuss that more Mn$^{3+}$ ions lead to higher activity, which is in line with the e$_g$ orbital descriptor proposed by Suntivich et al.[47]. In their catalytic model, comproportionation of Mn$^{4+}$ serves as a secondary supply of the catalytically active Mn$^{3+}$, which may explain that the lowest overpotential in systematic studies of the OER on simple manganese oxides is usually found for mixed Mn$^{3+/4+}$ oxides [3,4,38–40,42]. Therefore, it is critical to control the distribution of Mn cations in Mn-based electrocatalysts for the OER to understand their activity and potential degradation. Jaramillo and coworkers showed precious support materials (e.g. Au) induce Mn oxidation after the OER to such a mixed Mn$^{3+/4+}$ oxide [41,43]. Moreover, the voltage range in CV can have drastic influence on the observed redox features and thus the nature of the Mn oxide

[46,48]. Post-mortem and in situ experiments on many Mn oxides show clear valence changes during electrocatalytic experiments in alkaline media [37–40], often accompanied by structural changes [35,36]. Interestingly, studies including low voltages in addition to those supporting the OER often show the formation of $Mn_3O_4$ ($Mn^{2+}Mn^{3+}_2O_4$), which was not present in the pristine material [35,38,46]. The exact role of tetrahedral $Mn^{2+}$ in $Mn_3O_4$ is unknown but Rabe et al. [46] show that it hinders Mn dissolution using a combination of in situ Raman spectroscopy and ICP-MS.

In this study, we electrodeposited Na-containing $MnO_x$ films without long-rang order directly in an alkaline electrolyte using a complexing agent, while most other manganese oxide catalysts are deposited at lower pH (e.g. [12]). By this approach, we avoid complications due to ion exchange and focus on the changes of the Mn valence to identify activation and degradation processes during cyclic voltammetry and open-circuit conditions at pH 13. Electrochemical changes were observed and correlated with post-mortem XAS. Unexpectedly, we found that our $MnO_x$ films reactivate after open-circuit.

**Methods**

*Materials*

$Mn(NO_3)_2 \cdot 4H_2O$ (≥ 99.99 %), $MnO_2$ (≥ 99 %), $Mn_3O_4$ (≥ 97 %), $Mn_2O_3$ (≥ 99.9 %), L-(+)-Tartaric acid (≥ 99.5 %) and (2 M and 0.1 M) NaOH solutions were ordered from Sigma-Aldrich. Graphite foil (≥ 99.8) with a thickness of 0.254 mm ordered from VWR, deionized water (>18 MΩ cm). All reactants were used as received, without any further treatment.

*Electrodeposition of $MnO_x$ films*

0.6 mmol of $Mn(NO_3)_2 \cdot 4H_2O$ and 6 mmol of L-(+)-tartaric acid were dissolved in a small volume of deionized water (approx. 1 mL). 120 mL of Ar-purged 2 M NaOH solution were added slowly with stirring to the previous solution, changing from colorless to beige.

Electrodeposition was carried out using a Gamry 600+ potentiostat and a three-electrode cell made from a three-neck round-bottom flask. The separation between the necks and thus electrodes was less than 1 cm. The working electrodes were either a glassy carbon disk (4 mm diameter; HTW Sigradur G) in a rotating disk electrode (RDE) or graphite paper (Alfa Aesar). The unrotated RDE was mounted onto a commercial rotator (ALS RRDE-3A Ver 2.0). We used a saturated calomel reference electrode (SCE; ALS RE-2BP) and a graphite rod (redox.me, HP-III, High Pure Graphite) as the counter electrode. The galvanostatic deposition was performed at 150 µA/cm² until a charge density of 40 mC/cm² was reached.

*Electrochemical measurements*

The detailed protocol for electrocatalytic investigations is documented in Table S1 for glassy carbon electrodes and in Table S2 for graphite foil. The measurements on glassy carbon electrodes were carried out using two Gamry 600+ potentiostats connected as a bipotentiostat in a single-compartment three-electrode electrochemical cell made of polymethyl pentene (ALS) filled with about 60 mL solution of 0.1 M NaOH (pH 13). A commercial rotator (ALS RRDE3-A Ver 2.0) was used with commercial rotating ring-disk electrodes with exchangeable disks of 4 mm diameter and a Pt ring with inner ring diameter of 5 mm and outer diameter of 7 mm. The graphite foil was clamped in the same cell as the RRDE. A coiled platinum wire was used as a counter electrode and a SCE (ALS RE-2BP) as a reference electrode, which was calibrated daily against a commercial reversible hydrogen electrode (RHE; Gaskatel HydroFlex) as detailed in Fig. S1. The electrochemical experiments were performed at constant controlled temperature of 25.0 °C. The ring was set to detect oxygen at 0.4 V vs. RHE as calibrated previously [34]. Before any experiment, the electrolyte was purged with Ar for at least 30 minutes. Typical electrolyte resistances of R = 40±9 Ω were determined by electrochemical impedance spectroscopy (EIS) (representative data in Fig. S2a). The ohmic drop (also called iR drop) was corrected during post-processing by subtraction of iR from the measured voltages, where i and R are the measured current and resistance (Fig. S2b). All voltages are given relative to the reversible hydrogen electrode (RHE).

The instantaneous Tafel slope was calculated by the first derivative of the iR-corrected voltage as function of the logarithm of the current density. The Tafel slope was also calculated with a fitting of voltage as function of the logarithm of the current, using the cathodic half-cycle of the cyclic voltammetry of iR-corrected data in the range between 1.71 and 1.77 V vs. RHE. The electrodes were swept at 100 mV s$^{-1}$ and rotated at 1600 rpm. The Tafel slope was obtained by linear regression of the iR-corrected voltage (E-iR) against $\log_{10}(i)$. The error represents the standard deviation of three independently prepared electrodes.

### *Scanning electron microscopy (SEM) and energy dispersive X-ray spectroscopy (EDX)*

The sample morphology was determined using a Zeiss LEO Gemini 1530 scanning electron microscope, acceleration voltage of 3 keV and in high vacuum (around 10$^{-9}$ bar) and using a secondary electron inLens detector. The images were taken in different regions of the sample to get representative data. EDX measurements were performed using a Thermo Fischer detector with an acceleration voltage of 12 keV.

### *X-ray absorption spectroscopy (XAS)*

All XAS data were collected at an average nominal ring current of 300 mA in top-up and multi-bunch mode at the BESSY II synchrotron operated by Helmholtz-Zentrum Berlin.

Soft XAS measurements at the Mn-L edges were conducted using the LiXEdrom experimental station at the UE56/2 PGM-2 beamline [49]. Reference samples were measured as finely dispersed powders attached to carbon tape and electrodeposited samples were measured on graphite foil (Alfa Aesar). All samples were measured at room temperature and in total electron yield (TEY) mode and with horizontally linear polarization of the beam. The TEY measurements were carried out by collecting the drain current from the sample. The sample holder was connected to an ammeter (Keithley 6514). In order to avoid radiation damage, the incoming photon flux was adjusted to get a TEY current from the sample of around 10 pA. In addition, the sample was kept as thin as possible. XAS spectra for each sample were collected at a few locations to ensure representativity of the data and further minimize radiation damage and local heating. The energy axis was calibrated using a Mn-L edge spectrum of MnSO$_4$ as a standard where the maximum of the L$_3$-edge was calibrated to 641 eV (Fig. S3). This reference was calibrated against molecular oxygen as described elsewhere [38,50]. All spectra were normalized by the subtraction of a straight line obtained by fitting the data before the L$_3$ edge and division by a polynomial function obtained by fitting the data after the L$_2$ edge (Fig. S4).

Hard XAS at Mn-K edge was performed at the KMC-2 beamline [51]. The general used setup was organized as it follows: I$_0$ ionization chamber, sample, I$_1$ ionization chamber or FY detector, energy reference and I$_2$ ionization chamber. The used double monochromator consisted of two Ge-graded Si(111) crystal substrates[52] and the polarization of the beam was linear horizontal. Reference samples were prepared by dispersing a thin and homogenous layer of the powder on Kapton tape, after removing excess of powder, the tape was folded several times to get 2 cm x 1 cm windows. Reference samples were measured in transmission mode between two ion chambers detector at room temperature. Electrodeposited samples were measured on graphite foil in fluorescence mode with a Bruker X-Flash 6/60 detector. Energy calibration of the X-ray near edge structure (XANES) was made with the corresponding metal foil, setting the inflection point for Mn at 6539 eV. All spectra were normalized by the subtraction of a straight line obtained by fitting the data before the K edge and division by a polynomial function obtained by fitting the data after the K edge (Fig. S5). The Fourier transform (FT) of the extended X-ray absorption fine structure (EXAFS) was calculated between 40 and 440 eV (3.2 to 10.7 $A^{-1}$) above the Mn-K edge ($E_0$ = 6539 eV). A cosine window covering 10% on the left side and 10% on the right side of the EXAFS spectra was used to suppress the side lobes in the FTs.

EXAFS simulations were performed using the software SimXLite. After calculation of the phase functions with the FEFF8-Lite [53] program (version 8.5.3, self-consistent field option activated). Atomic coordinates of the FEFF input files were generated from the structure of MnOOH·xH$_2$O (birnessite) [54]; the EXAFS phase functions did not depend strongly on the details of the used model. An amplitude reduction factor (S0$^2$) of 0.7 was used. The data range used in the simulation was 44 to 420 eV (3.4 to 10.5 $A^{-1}$) above the Mn-K edge ($E_0$ = 6539 eV). The EXAFS simulations were optimized by the minimization of the error sum obtained by summation of the squared deviations between measured and simulated values (least-squares fit). The fit was performed using the Levenberg–Marquardt method with numerical derivatives.

**Results and Discussion**

*Electrodeposition of films on glassy carbon*

An electrodeposition protocol was developed to produce Mn oxides at alkaline pH. In these electrolytes, manganese might deposit in a numerous of stable oxide forms, commonly Mn$_3$O$_4$ (Mn$^{+2}$, Mn$^{3+}$), Mn$_2$O$_3$ (Mn$^{3+}$) or MnO$_2$(Mn$^{4+}$) [55]. Conventionally, solvated Mn$^{2+}$ cations would spontaneously precipitate as oxides or hydroxides at pH > 8. Therefore, we stabilized them using tartrate as a complexing agent to control their deposition in NaOH (inset of Fig. 1a). Furthermore, electrolyte anions may exchange with the deposited films, e.g. [31], which is prevented when the electrolyte composition does not change significantly during deposition and catalytic investigation.

Manganese oxides were deposited on glassy carbon electrodes for electrocatalytic investigations and on graphite paper for X-ray absorption spectroscopy (XAS). We first focus on the discussion of the samples deposited on glassy carbon disks. The deposition setup consisted of a commercial rotator for RDEs and a three-electrode cell (Fig S6). The films were deposited at constant current density of 150 µA/cm$^2$ (Fig. 1a), which allows to control the deposited charge by stopping the experiment after a certain time. The deposition profiles of MnO$_x$ showed increasing voltage after short times and reached a steady-state after about 20 s. The steady-state voltage of the samples was 1.66 ±0.04 V vs. RHE and the final deposited charge density after 34 s was 40 mC/cm$^2$.

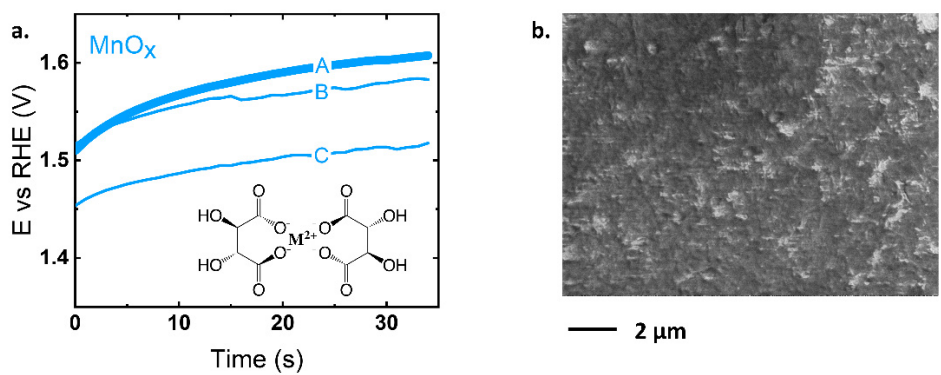

**Figure 1. a.** Chronoamperometry during electrodeposition of $MnO_x$ films on glassy carbon, samples A (thick line), B and C (thin lines). The inset shows the coordination complex of divalent metal ($M^{2+}$) due to the presence of tartrate ions. **b.** SEM image of sample A.

The film corresponding to $MnO_x$ sample A in Fig. 1a was subsequently characterized by SEM and EDX to check the coverage and homogeneity of the film. As it can been seen in Fig. 1b, $MnO_x$ covered the whole surface of the glassy carbon substrate with texture on the micrometer scale. EDX elemental mapping showed a homogeneous distribution of the electrodeposited Mn and O atoms (Fig. S8), thus corroborating the formation of $MnO_x$ anywhere on the substrate. Our films also contained Na (Fig. S8b), likely between layers of $MnO_x$. The crystallinity of the film was too low to be resolved by X-ray diffraction (XRD). In summary, the developed alkaline electrodeposition protocol ensured producing homogenously covered films of Na-containing $MnO_x$ without long-range order.

*Electrocatalytic investigations*

In order to evaluate the catalytic activity, activation and possible degradation of the films on glassy carbon disks, series of CV were performed in a RRDE station (representative data of $MnO_x$ sample A in Fig. 2 and all data in Fig. S7). The sweep speed was 100 mV s$^{-1}$, which is typical for activation studies, while cycling is a typical method for activation studies [30,56–58]. The CVs of the $MnO_x$ disks show an exponential increase of the current density at about 1.65 V vs. RHE for the 2$^{nd}$ cycle and 1.70 V vs. RHE for the 100$^{th}$ cycle. Additionally, about half the maximum current density was lost during 100 cycles. There were no additional redox peaks in the CV. The $MnO_x$ film disks showed hysteresis, i.e. capacitive currents, which reduced with cycling (insets of Fig. 2). Furthermore, the current offset reduced with cycling. The offset could be eliminated by selecting a wider scan range to more negative voltages (Fig. S9). The presence of the shifts in Fig. 2 may thus indicate an incomplete reduction process that is continued over many cycles.

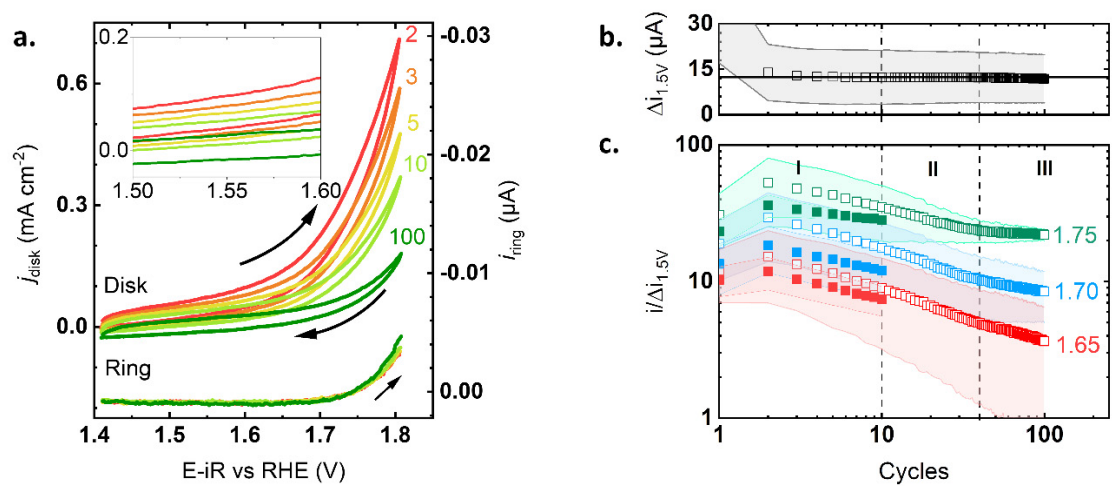

**Figure 2. a.** CV performed on a MnO$_x$-covered disk (sample A) with a scan rate of 100 mV s$^{-1}$ in 0.1 M NaOH with rotation 1600 rpm and constant voltage of 0.4 V vs. RHE at the ring. **b.** Average Δi$_{1.5V}$ of all samples as function of cycles for the first 100 cycles (calculation detailed in the text). **c.** Average current ratio i/Δi$_{1.5V}$ of all samples as function of cycling at selected voltages. The data was evaluated during the first 100 cycles (open squares) and 10 cycles after 30 minutes of OCV (solid squares). The dashed lines delimit three different regions, labeled as I, II and III.

The onset of the OER was determined using the anodic ring currents, which we assign to oxygen detection by reduction [34]. The cathodic ring scan is not shown as it shows hysteresis due to delayed oxygen release. The traces of the anodic ring currents were more similar than those of the disk currents. The onset of the OER on MnO$_x$ was consistently at 1.70 V vs. RHE independent of the cycle number as determined by the rise of the currents above the noise level. The differences in the trends between the disk and ring currents suggest that MnO$_x$ changes electrochemically with cycling without affecting the oxygen detected at the ring.

The trends of the disk currents during cycling were also evaluated at selected voltages with and without oxygen evolution (Fig. 2b,c). The currents were corrected for capacitance by averaging the anodic and cathodic scans [59]. To account for differences in surface area for the different films with the same composition, all currents were divided by the difference between the anodic and cathodic currents at 1.50 V vs. RHE (Δi$_{1.5V}$). After 5 cycles, Δi$_{1.5V}$ becomes steady with average values of 12±1 µA (Fig. 2b). The differential capacity is defined as C = dQ/dE ≈ Δi/(ΔE/Δt) where ΔE/Δt is the sweep speed (here: 100 mV s$^{-1}$). This is a rough approximation as the capacitance is commonly obtained by systematic experiments at several sweep speeds [60]. Yet, it allows tracking the changes of surface area with cycling using the regular CV method. SEM showed no significant morphology changes after 100 cycles for the three MnO$_x$ samples (Fig. S10).

The current ratios *i*/Δ*i*$_{1.5V}$ of MnO$_x$ followed a power law with different exponents depending number of cycles and whether oxygen is evolved (Fig. 2c; open symbols). A large negative exponent means fast decay of the current with cycle number, while a large positive exponent means a fast increase of the currents with cycling. The exponent was about -1/3 for the initial 10 cycles (region I), about -1/4 for intermediate cycles (10 to 40; region II) and depended on the applied voltage for the later cycles, i.e. after about 40 cycles (region III). The exponent was similar to the initial value of -1/3 before the onset of the OER (1.65 V vs. RHE); it was about -1/5 at the onset (1.70 V) and close to zero for oxygen evolution (1.75 V vs. RHE). The spread of the currents (light colored area) was smallest for the steady-state (zero exponent) at this voltage.

In summary, the currents decreased with cycling for $MnO_x$ and reached a steady-state only at the voltage which support oxygen evolution (1.75 V vs. RHE). These trends can be explained by (1) change in coverage with intermediates (hydroxylation or $O_2$ coverage) [1,61], (2) material dissolution [17], modifications of the catalyst material such (3) change in structure [19] or (4) transition metal valence [18]. The transient changes of point (1) can only be investigated in situ, while the irreversible changes of points (2)-(4) are expected to persist in a post-mortem experiment. Other explanations may be possible, but we deem these points the most likely ones and thus we address these possible explanations of the trends point by point below.

To distinguish irreversible surface changes and transient changes, we let the films rest at open-circuit voltage (OCV) for 30 minutes and then performed an additional 10 cycles. This simple test was recently introduced by El-Sayed et al. [62] for galvanostatic measurements using RDEs of iridium on ATO (antimony-doped tin oxide). The authors attribute the unexpected recovery of currents after OCV to the removal of nano and micro bubbles of oxygen within the catalyst layer, i.e. a change of (product) coverage. Here, we extend it to electrocatalyst tests using CV (Fig. 2b; solid symbols).

The exponents of the current ratio differed depending on the previously applied voltage after the OCV period for all films, suggesting partially reversible processes. The exponent was identical (-1/3) below the onset of the OER and less negative at the onset (-1/4) and above (-3/20). Furthermore, the current ratios were offset to lower values, i.e. the prefactor of the power law was lower. We interpret these observations as follows: the reduced prefactor indicates irreversible changes due to a reduction of active sites, either by material dissolution or deactivation of the active sites. Note that we normalized by a quantity proportional to the surface area so that we can exclude it as the source of the lower prefactor and hence currents. We cannot exclude dissolution of specific cations that leads to a different surface composition as observed, e.g. for the perovskite $La_{0.6}Sr_{0.4}MnO_3$ [63]. We attribute the changes in the exponent to a combination of transient changes of species adsorbed on the surface and the irreversible changes listed above.

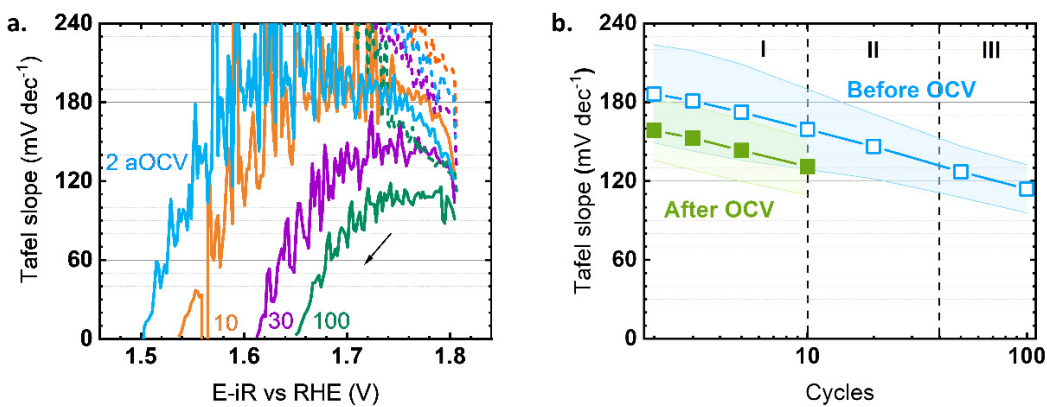

**Figure 3. a.** Instantaneous Tafel slope as function of voltage of cycles 10, 30 and 100 before OCV and cycle 2 after OCV (2aOCV). All Tafel slope values were calculated from CVs with a scan rate of 100 mV s$^{-1}$ in 0.1 M NaOH rotated at 1600 rpm. Arrows show the direction of the scan. **b.** Averaged Tafel slope as function of cycle before (open symbols) and after (solid symbols) OCV. The light-colored areas represent the standard deviation of three samples. The Tafel slopes were calculated between 1.73 V – 1.78 V vs. RHE.

The Tafel slope ($b = \partial \log i / \partial E$) in chemical equilibrium depends on the surface coverage where usually limiting cases of the adsorption isotherms are discussed [1,61,64,65]. Since these values do not depend on the amount of compound or active sites, they are usually associated with mechanisms of specific reactions and vast reported values can be found for a wide range of materials in OER catalysis. The Tafel slope is very large (approaching infinity) if an early chemical step limits the reaction without any pre-equilibria [61]. In particular, very large Tafel slopes are also expected if the intermediate that reacts chemically during the rate-limiting step approaches full surface coverage in the Langmuir adsorption model [61]. A Tafel slope of 120 mV dec$^{-1}$ is predicted (at 25 °C) if an electrochemical step limits the kinetics, if the electrochemically reacting intermediate reaches full coverage in the Langmuir model or if the chemically reacting intermediate reaches full coverage in the Temkin model [61]. A Tafel slope of 60 mV dec$^{-1}$ (at 25 °C) indicates a chemical rate-limiting step with an electrochemical pre-equilibrium for low coverage of the reacting intermediate (in the Langmuir model) [1,61]. Values between 60 and 120 mV dec$^{-1}$ are not predicted by common kinetic models. Thus, the Tafel slope does depend on the surface coverage but not uniquely.

We analyzed the Tafel slope during selected cycles of a representative sample (Fig. 3, all samples in Fig. S11). The Tafel slopes during the anodic scans in Fig. 3a were larger as compared to those of the cathodic scans. Clearly a sweep speed of 100 mV s$^{-1}$ was too fast to establish a chemical equilibrium. If an electrochemical equilibrium plays a mechanistic role, then the produced intermediates were biased toward the oxidative site of the equilibrium (e.g. M$^{III}$OH → M$^{IV}$O) during the cathodic scan, due to the voltages above the equilibrium voltage. During the 10$^{th}$ scan, the Tafel slope of MnO$_x$ was 180 mV dec$^{-1}$ at the onset of the OER and below it (1.7 to 1.6 V vs. RHE). Note that previously electrochemically produced intermediates can react chemically below the onset of the OER (as determined by the ring during the anodic scan). After 100 cycles, the Tafel slope decreased to 110 mV dec$^{-1}$ between 1.8 and 1.7 V vs. RHE. In fact, the Tafel slopes decreased exponentially before and after the OCV (Fig. 3b and S12). After the OCV (2aOCV), the trace of the Tafel slope and its constant value of 180 mV dec$^{-1}$ matched that of the 10$^{th}$ cycle before the OCV (Fig. 3a). In summary, the region of constant Tafel slope (or minimum) shifted to lower voltages with cycling for both materials. The Tafel slope decreased for MnO$_x$ with cycling. After OCV, the voltage dependence and value (within error) of the Tafel slope matched those observed initially. We interpret these trends as a reversible formation of an active state (with certain intermediate coverage) on MnO$_x$.

### *Local structure and valence changes due to cycling*

We used XAS to study the irreversible changes due to cycling. XAS experiments were necessary due to its fine chemical sensitivity and the absence of crystallinity required for diffraction techniques. Changes in the local structure of the oxides were identified using the EXAFS at the Mn-K edge [66–68]. The XANES was used to discuss changes of the metal valence where we analyzed the bulk-sensitive K edge and more surface-sensitive L edge of Mn.

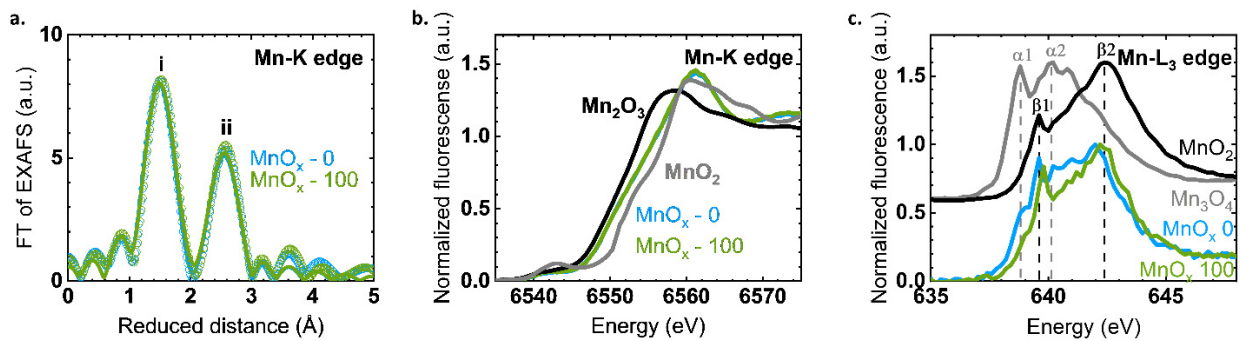

**Figure 4. a.** Fourier-transform EXAFS spectra for Mn K-edge collected on pristine $MnO_x$ (blue symbols) and after 100 cycles of OER catalysis (green symbols). The corresponding lines are results from EXAFS simulations (see Table 1 for parameters). The reduced distance is by about 0.3 Å shorter than the precise distance obtained by EXAFS simulations. **b.** XANES spectra for Mn-K edge collected on $MnO_x$ after 0 cycles and 100 cycles. The Mn-K edge spectra of $Mn_2O_3$ and $MnO_2$ were added as references. **c.** Mn-$L_3$ edge spectra for pristine $MnO_x$ ($MnO_x$ 0) and $MnO_x$ after 100 cycles ($MnO_x$ 100); and $MnO_2$ and $Mn_3O_4$, which were used as references.

The sample for the XAS investigations were prepared on graphite foil by the same procedure as discussed above (Table S2). The deposition of the same charge density on the larger graphite foil also resulted in a steady-state after about 20 s but at lower voltages of 1.37±0.03 V vs. RHE (Fig. S7). The systematic shift by about +300 mV did not affect the observed electrocatalytic behavior with cycling (Fig. S14). While there may have been minor variations of the film composition, valence and roughness due to the different steady-state voltages, the identical cycling trends strongly suggest that the trends in the XAS data can be used to rationalize the cycling trends of films deposited on both graphite foil and glassy carbon.

**Table 1.** EXAFS absorber–backscatter distance (R), coordination numbers (N) and Debye–Waller factor (σ) as determined by simulation of the $k^3$-weighted EXAFS spectra at the Mn-K edge for pristine $MnO_x$. ($MnO_x$-0) and after 100 cycles ($MnO_x$-100). Shells were simulated using phase functions from a previously reported birnessite structure [54].

| Sample | Parameter | Mn–O1 | Mn–O2 | Mn–Mn | R – factor |
|---|---|---|---|---|---|
| **$MnO_x$-0** | N | 5.0 | 1.0 | 4.3 | |
| | R (Å) | 1.87 | 2.30 | 2.86 | 0.59 % |
| | σ (Å) | 0.05 | 0.05* | 0.08 | |
| **$MnO_x$-100** | N | 5.2 | 1.5 | 4.2 | |
| | R (Å) | 1.87 | 2.30 | 2.86 | 0.85 % |
| | σ (Å) | 0.05 | 0.05* | 0.08 | |

* indicates fixed values (not simulated).

The R-factor used Fourier filtered data between 1 and 3 Å using the formula $Rf = 100\ \Sigma(m_i^{ff}-e_i^{ff})^2/\Sigma(e_i^{ff})^2$, where $m^{ff}$ represents the Fourier-filtered model and $e^{ff}$ represents the experimental k-weighted EXAFS curve.

The Fourier transform of the EXAFS of $MnO_x$ showed the expected features of layered hydroxides (Fig. 4a), namely two peaks at about 1.5 and 2.5 Å reduced distance labeled i and ii, respectively. The FT of the EXAFS did not change with cycling for these peaks. We simulated peaks i and ii using $MnOOH \cdot xH_2O$ (birnessite) [54] (Fig. 4a). The fit results corroborate the assignment to a layered hydroxide as the R-factors were below 2 % [68] (Table 1). There were no drastic changes with cycling in the local structure of the bulk but we cannot exclude surface changes (additional discussion below). The metal-oxygen bond length is 1.87 Å (peak i), which is typical for $Mn^{3+/4+}O_6$ cations [54,69]. Yet, there is another Mn-O bond length of around 2.30 Å, which significantly improves the fit. This bond length is similar to the one in $Mn^{2+}O$ and may indicate a minor impurity phase [54]. The Mn-Mn distances were 2.87 Å in $MnO_x$. These distances have been widely reported for amorphous electrodeposited oxides where they correspond to hexacoordinated Mn-O and di-μ-oxo bridged metals, i.e. edge-sharing octahedra [3,70]. The absence of clear FT peaks at higher reduced distance implies a lack of long-range order.

A qualitative estimation of the average bulk metal valence can be achieved by comparison of the XANES spectra of the films and suitable reference materials (Fig. 4b). The Mn-K edge spectra of $MnO_x$ were compared with the two references $Mn^{3+}_2O_3$ and $\beta\text{-}Mn^{4+}O_2$. The edge rise of our films falls between that of the references, which indicates that the average Mn valence is between +3 and +4 in the film. The spectra of $MnO_x$ showed negligible changes.

As catalysis is a surface process, we turned our attention to soft X-ray spectroscopy accessed by the electron yield mode. The escape depth of the electrons is 2.6±0.3 nm at the Mn-L edge of a comparable oxide [71], so that we probe the near surface regions of the samples. In contrast to the K edge XANES, we found clear changes with cycling near the surface at the $L_3$ edge (Fig. 4c). The main peaks of the references $Mn^{2.6+}_3O_4$ and $\beta\text{-}Mn^{4+}O_2$ were labeled $\alpha 1$-$\alpha 2$ and $\beta 1$-$\beta 2$, respectively. Peak $\alpha 1$ can be assigned to $Mn^{2+}$ in tetrahedral coordination [38]. The Mn-$L_3$ spectrum of pristine $MnO_x$ was broad and contained features of both references. We found a clear reduction of the normalized spectral intensity at the energies assigned to peaks $\alpha 1$ and $\alpha 2$ after 100 cycles (Fig. 4b). The spectrum of $MnO_x$ after 100 cycles closely resembled that of the $MnO_2$ reference but a slight shoulder at the energy of $\alpha 1$ remained.

We conclude that the $MnO_x$ films were deposited as layers of edge-sharing octahedra with low long-range order. The local bulk structure was preserved upon cycling. There were no changes in the bulk, while the surface region of $MnO_x$ oxidizes toward $Mn^{4+}$ and $Mn_3O_4$ or another phase that contained tetrahedral $Mn^{2+}$ was consumed but did not vanish completely.

Having uncovered the local structure and surface valence changes on $MnO_x$, we can now interpret the trends during electrocatalytic cycling in this context. Irreversivle catalyst changes lead to a decrease of the current ratio at all voltages below the onset of the OER. The current ratio at low overpotential achieved steady-state after 50 cycles, which did not persist after OCV. This irreversbile decrease in current ratio could either be explained by loss of tetrahedral $Mn^{2+}$, possibly as $Mn_3O_4$, at the surface or surface oxidation toward $Mn^{4+}$. While the role of the former is not well understood, there is a broad consensus on the role of $Mn^{4+}$ as discussed in the next paragraphs.

The most commonly studied oxide with tetrahedral $Mn^{2+}$ is $Mn_3O_4$, which consists of tetrahedral $Mn^{2+}$ and octahedral $Mn^{3+}$ cations. Some studies state that (bulk) $Mn_3O_4$ is an active electrocatalyst [12,72–74], while it is the least active catalyst in comparative studies [11,25]. Thus, it is unclear if bulk $Mn_3O_4$ is an electrocatalyst as its electric conductivity is low [35] and other Mn oxide phases are found at the surface of bulk $Mn_3O_4$ such as amorphous $Mn^{3+/4+}$-oxide [75] or birnessite-type $MnO_x$ [11,76], both of which are formed at voltages below the onset of the OER [77,78]. Huynh et al.[79] identified $Mn_3O_4$ as a precursor to an active birnessite-like phase. Finally, Wei et al.[80] propose that only the octahedral Mn site is active for the OER in spinel oxides. In summary, tetrahedral $Mn^{2+}$ is likely not directly relevant during the OER but may be important for the formation of the active phase and as a passivating phase that prevents Mn leaching [46].

The role of (octahedral) $Mn^{4+}$ for the OER is better understood. $Mn^{4+}$ forms at voltages below the onset of the OER [38,46]. While many of the more active Mn-based electrocatalysts contain $Mn^{4+}$ in addition to $Mn^{3+}$ [10,39,40,81–84], those with predominately $Mn^{4+}$ are inactive [3]. Recently, Baumung et al. [18] showed that (bulk) oxidation of $LiMn^{3.5+}{}_2O_4$ toward $Mn^{4+}$ increases the overpotential of the OER. This can be rationalized qualitatively using the occupancy of the $e_g$ orbitals as proposed by Suntivich et al. [47] where the lowest overpotential is predicted for an $e_g$ occupancy of about one ($Mn^{3+}$) and both higher or lower $e_g$ occupancy (i.e. $Mn^{2+}$ and $Mn^{4+}$) leads to an increase in overpotential. Yet, small amounts of $Mn^{2+}$ and $Mn^{4+}$ are beneficial if Mn comproportionation of these ions to $Mn^{3+}$ is possible on the surface of the Mn oxide [44,45]. Thus, the lowest overpotential in systematic studies of the OER on manganese oxides is usually found between $Mn^{3.5+}$ and $Mn^{3.7+}$ [3,4,38–40,42], rather than $Mn^{3+}$. Nonetheless, materials with mainly $Mn^{4+}$ perform significantly worse than those with sizable amount of $Mn^{3+}$, e.g. Köhlbach et al. [85] attributed the formation of $Mn^{4+}$ on the surface to deactivation of $\alpha$-$Mn_2O_3$ films and Rabe et al. [46] identified the highest rate of Mn dissolution on $MnO_2$ during CV. Therefore, we conclude that the $Mn^{4+}$ on the surface of our $MnO_x$ films is responsible for the observed decrease in the current. Future studies should address the optimal amount of $Mn^{4+}$ at the surface during the oxidizing conditions of the OER to maximize activity and to simultaneously prevent deactivation processes during catalyst conditioning or operation.

In addition to the previously reported irreversible change of manganese oxides, we also found that the film could be partially reactivated after 30 minutes at OCV (Fig. 2b and Fig. 3). The OCV was 0.97 V or lower and reached a steady-state for one sample at 0.88 V vs. RHE (Fig. S16). At these voltages, reduction of $Mn^{4+}$ to $Mn^{3+}$ is expected as shown by an in situ soft XAS study [38] and chemical reactions such as oxygen evolution by reduction of $Mn^{4+}$ may occur [3]. We interpreted the increase of the measured current and of the Tafel slope as the reversible formation of an active state with a certain, still unknown, high coverage of catalytic intermediates, which could be associated with a valence change of the $MnO_x$ surface. The elucidation of possible valence changes and thus the active state requires specialized in situ experiments which are beyond the scope of this report. We note that the identical exponents initially and after OCV suggest that $Mn^{4+}$ is produced at the same rate after reactivation. The reactivation after OCV is significant for catalysis on manganese oxides and their activation protocol. Degradation tests based purely on electrochemical methods such as CV or voltage/current holding may suggest an irreversible material degradation while an optimized measurement protocol could recover some activity.

## Conclusions

We deposited a Na-containing layered oxide of $Mn^{3+/4+}$ without long-range order in NaOH solution with a complexing agent added. The onset of the OER was at 1.7 V vs. RHE as determined by oxygen detection at the ring of an RRDE. We tracked the currents by CV during 100 cycles where 3 regions with different trends could be identified: 1-10 cycles, 10-40 cycles and 40-100 cycles. At voltages below the onset of the OER, the currents decrease with cycling but the OER currents at 1.75 V vs. RHE reached a steady-state in the 3$^{rd}$ region. Thus, an activation protocol at 100 mV s$^{-1}$ should at least be performed for 40 cycles on our $MnO_x$. The bulk of the cycled $MnO_x$ film was not changed but its surface was oxidized toward $Mn^{4+}$. It agrees with previous reports that the oxidized $MnO_x$ surface hinders the OER. Interestingly, the currents during CV could be partially recovered after 30 minutes of OCV, which was not observed previously. We attribute the underlying process to high coverage with unreacted intermediates as supported by the high Tafel slopes. The reactivation after OCV is significant as some manganese oxides or Mn-containing oxides may be more robust as estimated based on continuous cycling. A measurement protocol without continuous cycling may prevent some but not all activity-reducing processes. Coupling of electrocatalysts to intermittent sources such as solar energy naturally leads to OCV conditions during operation. Therefore, it is important how an electrocatalyst for sustainable fuel production reacts to OCV, which is understudied.

## Acknowledgements

We acknowledge Denis Antipin, Max Baumung, Florian Schönewald, Dr. Daowei Gao and Dr. Laura Pardo for helping in data collection. Frederik Stender is thanked for writing the electrochemistry analysis script and Dr. Petko Chernev for permission to use his software SimXLite. We thank Helmholtz-Zentrum Berlin (HZB) for the allocation of synchrotron radiation beamtime and acknowledge HZB CoreLab Correlative Microscopy and Spectroscopy for training and advising in SEM. This project has received funding from the European Research Council (ERC) under the European Union's Horizon 2020 research and innovation programme under grant agreement No 804092.

## ORCID IDs

Javier Villalobos (0000-0002-8032-6574)

Götz Schuck (0000-0002-0624-2719)

Lifei Xi (0000-0002-5248-7936)

Marcel Risch (0000-0003-2820-7006)

## Conflict of interest

The authors declare no conflict of interest.

SUPPLEMENTARY MATERIAL

# Reversible and irreversible processes during cyclic voltammetry of an electrodeposited manganese oxide as catalyst for the oxygen evolution reaction

Javier Villalobos[1], Ronny Golnak[2], Lifei Xi[1], Götz Schuck[3] and Marcel Risch[1,4],*

[1] *Nachwuchsgruppe Gestaltung des Sauerstoffentwicklungsmechanismus, Helmholtz-Zentrum Berlin für Materialien und Energie GmbH, Hahn-Meitner Platz 1, 14109 Berlin, Germany*

[2] *Department of Highly Sensitive X-ray Spectroscopy, Helmholtz-Zentrum Berlin für Materialien und Energie, Albert-Einstein-Straße 15, 12489 Berlin, Germany*

[3] *Department of Structure and Dynamics of Energy Materials, Helmholtz-Zentrum Berlin für Materialien und Energie, Hahn-Meitner Platz 1, 14109 Berlin, Germany*

[4] *Institut für Materialphysik, Georg-August-Universität Göttingen, Friedrich-Hund-Platz 1, 37077 Göttingen, Germany*

**Corresponding author**

* marcel.risch@helmholtz-berlin.de

11 pages, 3 supplementary tables, 16 supplementary figures

**Table S1.** General protocol for electrochemical data collection on an RRDE station with samples on glassy carbon disks. All potentials are reported vs. RHE (pH 13).

| Step | Conditions |
|---|---|
| **1.** Cleaning | Clean and polish electrodes, cells and any other tool properly. |
| **2.** Calibration of reference electrodes | OCV against commercial RHE electrode |
| **3.** Argon purge at OCV | At least 30 minutes |
| **4.a.** Ring EIS | Frequency: 1 MHz – 1 Hz.<br>Points/decade: 10<br>OCV and take note of Ru |
| **4.b.** Disk EIS | Frequency: 1 MHz – 1 Hz.<br>Points/decade: 10<br>OCV and take note of Ru |
| **5.** Disk CV: ECSA* | Hold 10 s at 1.0 V.<br>Potential window: 0.95 V – 1.05 V.<br>Scan rates: 50, 100, 150, 200, 250, 300, 350, 400, 450, 500 mV s$^{-1}$.<br>Cycles: 3.<br>Rotation: 0 rpm.<br>Purge: Blanket.<br>No dynamic iR compensation. |
| **6.a.** Ring conditioning | Hold ring potential for 1800 s at 0.40 V. |
| **6.b.** Ring: CA ($O_2$ detection) | Hold ring potential at 0.40 V. |
| **6.c.** Disk CV: OER | Potential window: 1.40 V - 1.80 V<br>Scan rate: 100 mV/s<br>Step size: 2 mV<br>Cycles: 100<br>Rotation: 1600 rpm<br>Purge: yes<br>No dynamic iR compensation. |
| **7.** Disk CV: ECSA* | Hold 10 s at 1.0 V.<br>Potential window: 0.95 V – 1.05 V.<br>Scan rates: 50, 100, 150, 200, 250, 300, 350, 400, 450, 500 mV s$^{-1}$.<br>Cycles: 3.<br>Rotation: 0 rpm.<br>Purge: Blanket.<br>No dynamic iR compensation. |
| **8.** Disk OCV | 1800 s |
| **9.a.** Disk EIS | Frequency: 1 MHz – 1 Hz.<br>OCV and take note of Ru |
| **9.b.** Ring EIS | Frequency: 1 MHz – 1 Hz.<br>OCV and take note of Ru |
| **10.a.** Ring conditioning | Hold ring potential for 1800 s at 0.40 V. |
| **10.b.** Ring: CA ($O_2$ detection) | Hold ring potential at 0.40 V. |
| **11.** Disk CV: OER | Potential window: 1.40 V - 1.80 V<br>Scan rate: 100 mV/s<br>Step size: 2 mV<br>Cycles: 10<br>Rotation: 1600 rpm<br>Purge: yes<br>No dynamic iR compensation. |

\* not used due to inappropriate data for analysis**.**

**Table S2.** General protocol for electrochemical data collection with samples on graphite foil. All potentials are reported vs. RHE (pH 13).

| | |
|---|---|
| **1.** Cleaning | Clean and polish electrodes, cells and any other tool properly. |
| **2.** Calibration of reference electrodes | OCV against commercial RHE electrode |
| **3.** Argon purge at OCV | At least 30 minutes |
| **4.** Foil CV: OER | Potential window: 1.40 V - 1.80 V<br>Scan rate: 100 mV/s<br>Step size: 2 mV<br>Cycles: 100<br>Rotation: 1600 rpm<br>Purge: yes<br>No dynamic iR compensation |
| **5.** Sample rinsing | Soaked in deionized water for 5 minutes. |

**Table S3.** Summary of the fit parameters from three sets of data used for Tafel slope calculation and their average. One example of the graphs involved in the calculation is shown in Figure S11. The average of the three values for each cycle and each sample is used to generate Figure 3b. SD represents the standard deviation of the three measurements.

| Cycle | Slope ± error ($R^2$) | | | Average ± SD |
|---|---|---|---|---|
| | **Sample A** | **Sample B** | **Sample C** | |
| 2 | 170 ± 1 (0.999) | 160 ± 1 (0.999) | 230 ± 1 (0.999) | 186 ± 48 |
| 3 | 168 ± 1 (0.999) | 151 ± 1 (0.999) | 224 ± 1 (0.999) | 181 ± 46 |
| 5 | 165 ± 1 (0.999) | 141 ± 1 (0.999) | 212 ± 1 (0.999) | 172 ± 37 |
| 10 | 162 ± 1 (0.999) | 128 ± 1 (0.999) | 189 ± 1 (0.999) | 159 ± 30 |
| 20 | 159 ± 1 (0.999) | 118 ± 1 (0.999) | 162 ± 1 (0.999) | 146 ± 25 |
| 50 | 147 ± 1 (0.999) | 108 ± 1 (0.999) | 126 ± 1 (0.999) | 127 ± 20 |
| 100 | 135 ± 1 (0.999) | 102 ± 1 (0.998) | 106 ± 1 (0.999) | 114 ± 18 |
| 2aOCV* | 165 ± 1 (0.999) | 133 ± 1 (0.999) | 179 ± 1 (0.999) | 159 ± 23 |
| 10aOCV* | 149 ± 1 (0.999) | 107 ± 1 (0.999) | 137 ± 1 (0.999) | 131 ± 22 |

* aOCV = after OCV break.

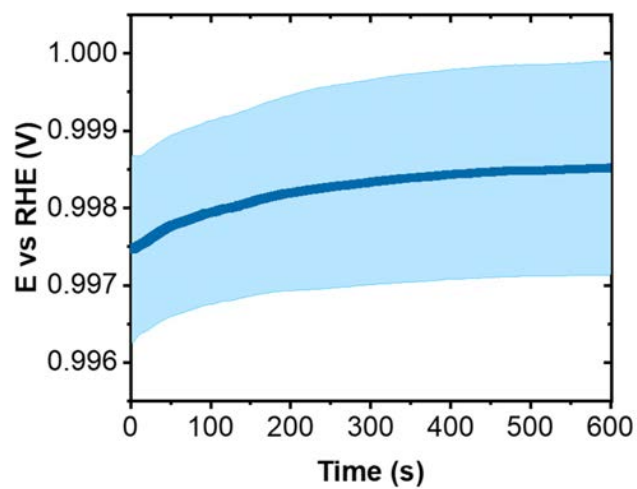

**Figure S1.** Calibration of reference electrode (SCE) vs. RHE. The blue line represents one single measurement and the light blue shade the error (standard deviation) of the measurements of the same two electrodes on three different days.

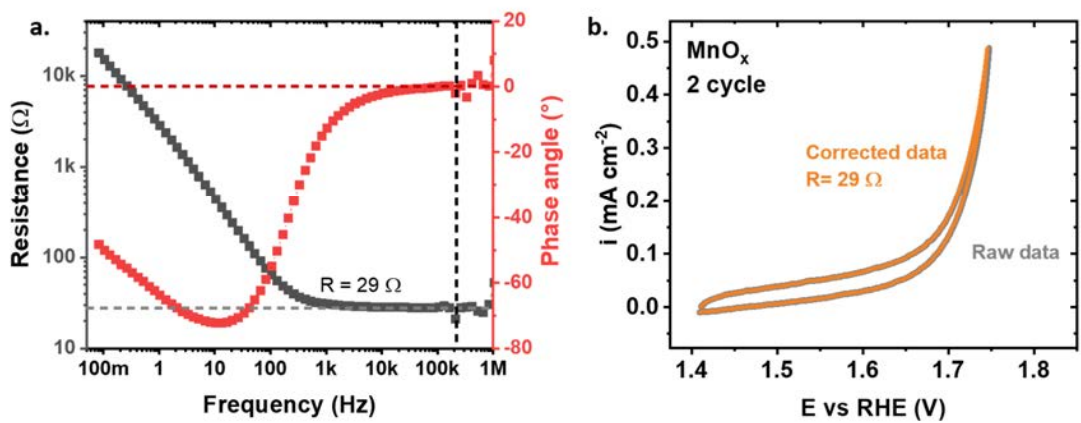

**Figure S2.** Example of data treatment of $MnO_x$ after 2 cycles of catalysis: **a.** Electrochemical impedance spectrum (EIS) showing resistance (black) and phase angle (red) as function of frequency. Dashed lines represent the resistance value where the phase angle approaches zero. **b.** CV of cycle 2 with the correction of the raw data (gray) by a resistance value = 29 Ω, and the resulting corrected CV (orange).

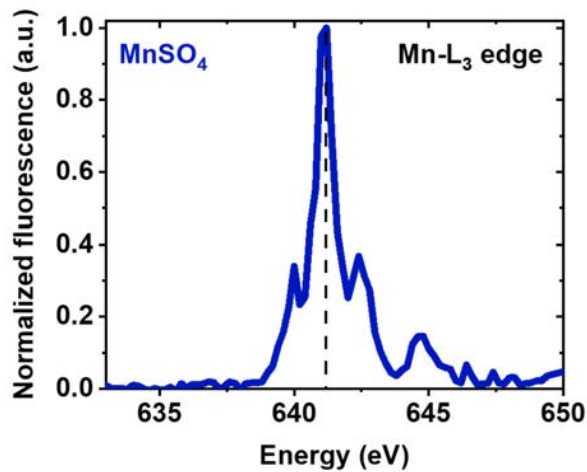

**Figure S3.** Mn-L$_3$ edge spectrum of MnSO$_4$, used as a reference for energy calibration. The dashed line was calibrated to 641 eV.

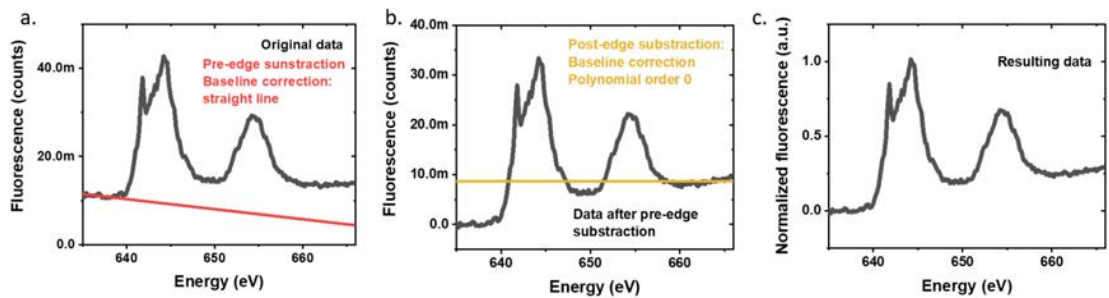

**Figure S4.** Spectra of a representative normalization for soft XAS of Mn-L$_3$ edge: **a.** Raw data (black) and baseline (red) for fitting before the edge (fit between 632 and 638 eV), **b.** Data after pre-edge subtraction (black) and baseline for fitting after the edge (yellow; fit between 660 and 667 eV ), **c.** Final normalized spectrum.

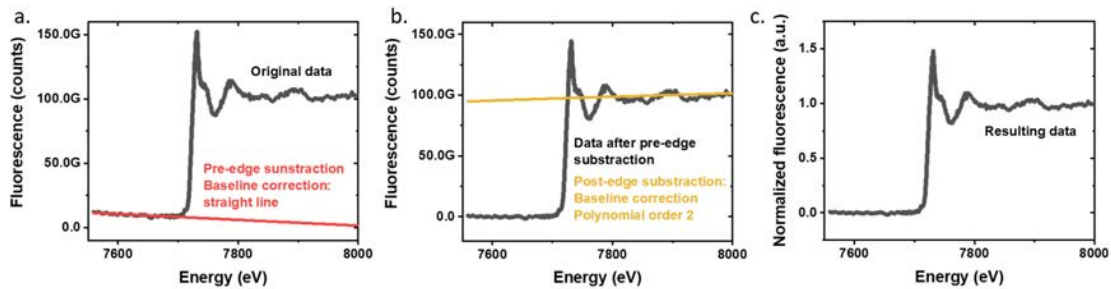

**Figure S5.** Spectra of a representative normalization for hard XAS of Mn-K edge (zoom on near edge region): **a.** Raw data (black) and baseline (red) for fitting before the edge (fit between 7560 and 7695 eV), **b.** Data after pre-edge subtraction (black) and fitted polynomial for division after the edge (yellow; fit between 7960 and 8255 eV) on a wider energy range as the shown spectrum, **c.** Final normalized spectrum.

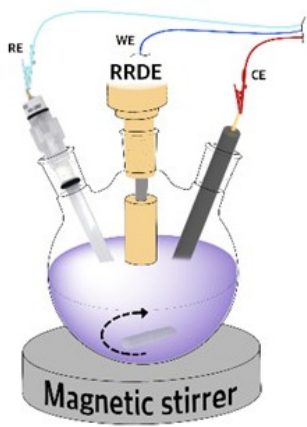

**Figure S6.** Scheme of the 3-electrode cell and RDE used for the electrodeposition of the films on glassy carbon.

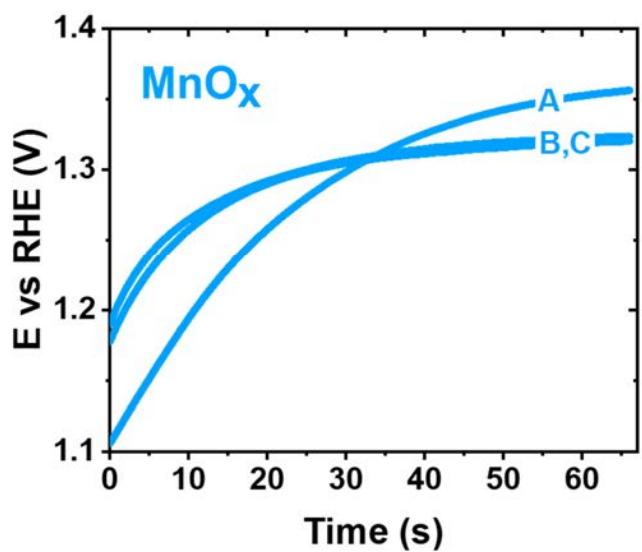

**Figure S7.** Chronopotentiometry during eletrodeposition of MnO$_x$ samples A, B and C on graphite foil to be characterized by XAS.

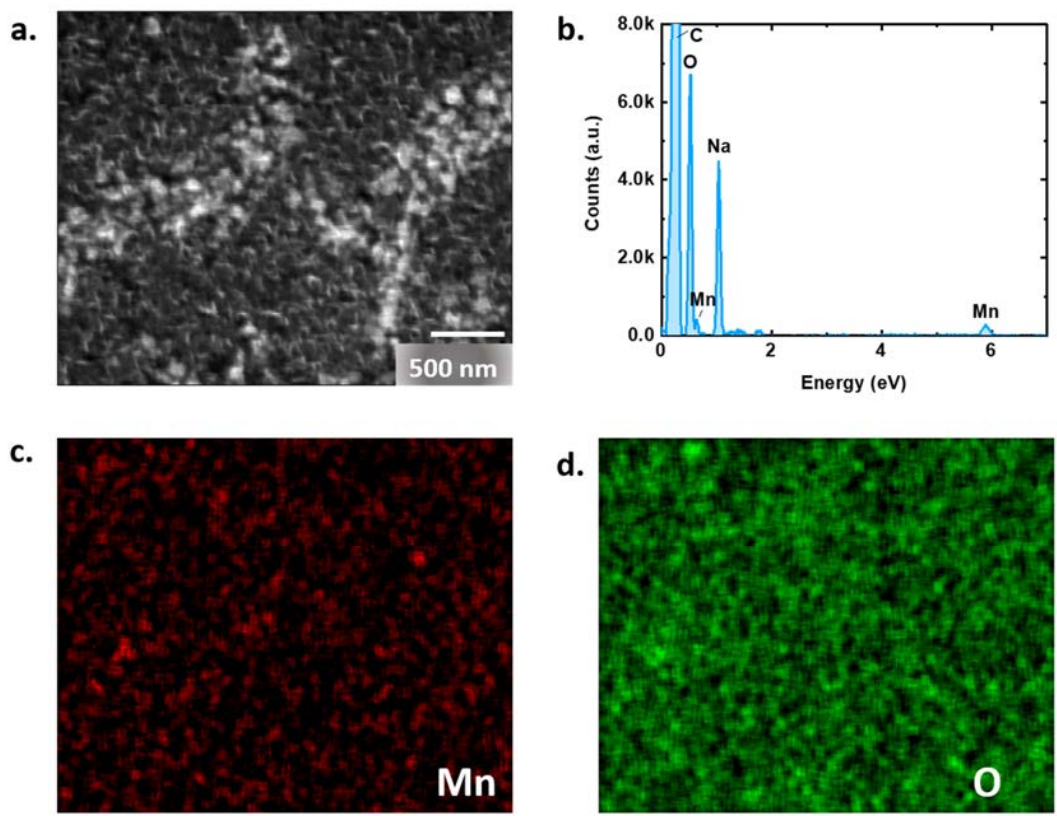

**Figure S8.** SEM-EDX analysis of pristine MnOx films on glassy carbon. **a.** Gray contrast image, **b.** EDX spectrum. Elemental mapping by EDX of **c.** manganese and d. oxygen.

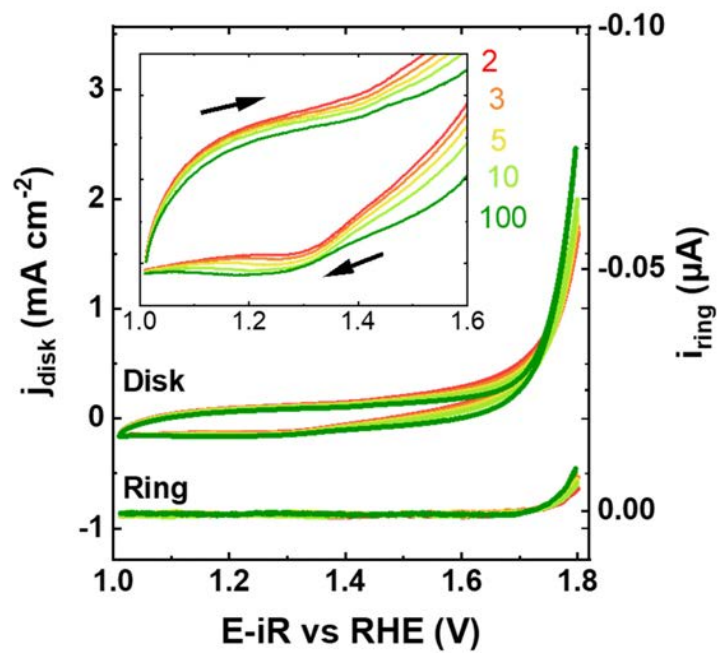

**Figure S9.** CV performed on a MnO$_x$-covered disk and step potential of 0.4 V vs. RHE at the ring. The was performed with a scan rate of 100 mV s$^{-1}$ in 0.1 M NaOH with rotation: 1600 rpm. This sample was measured with lower voltage boundary as samples shown in Fig. 2a. The inset shows a zoom of the region of interest due to the presence of redox peaks. The arrows indicate the direction of the scan.

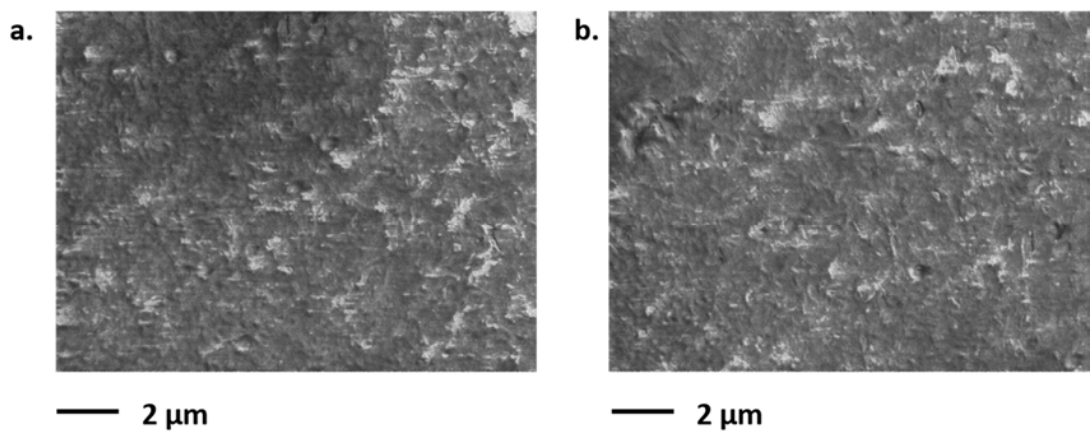

**Figure S10.** SEM images of: **a.** Pristine MnO$_x$ and **b.** MnO$_x$ after 100 cycles of OER catalysis. These measurements correspond to sample A.

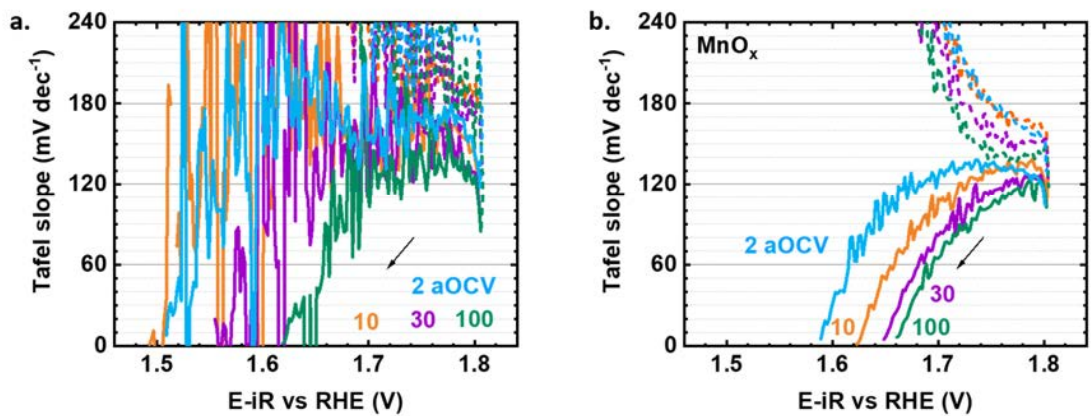

**Figure S11.** Instantaneous Tafel slope as function of potential of additional $MnO_x$ films: **a.** Sample A and **b.** Sample B. Before OCV, cycles 10, 50 and 100 are shown, and after OCP cycle 2 is shown (2aOCV). All Tafel slope values were calculated from CVs with a scan rate of 100 mV s$^{-1}$ in 0.1 M NaOH rotated at 1600 rpm. iR correction was done in post-processing. Arrows shows the direction of the CV.

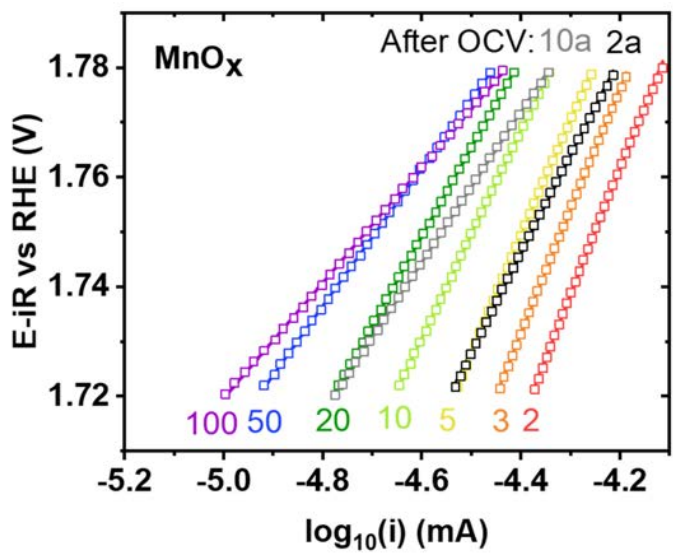

**Figure S12.** Representative graphs of Tafel slope calculation for selected cycles (2, 3, 5, 10, 20, 50, 100, 2 after OCV and 10 after OCV). Tafel plots of $MnO_x$. The measurements were performed in 0.1 M NaOH pH 13. Scan rate 100 mV s$^{-1}$. iR compensation was performed during post-processing. The lines represent the linear fit (denoted E-iR) of E-iR as function of $\log_{10}(i)$, the slope represents the Tafel slope. All fit parameters are summarized in Table S3.

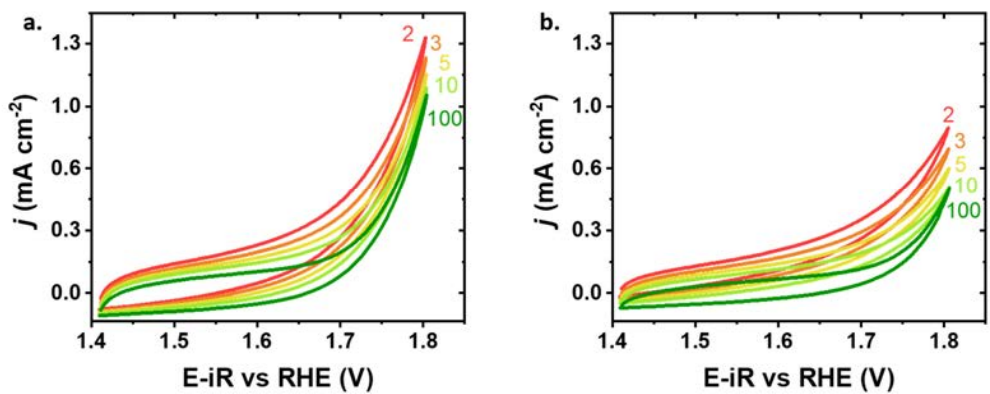

**Figure S13.** Series of CV of additional MnO$_x$ films: **a.** Sample B and **b.** Sample C, performed at the disk. CV with a scan rate of 100 mV s$^{-1}$ in 0.1 M NaOH rotated at 1600 rpm.

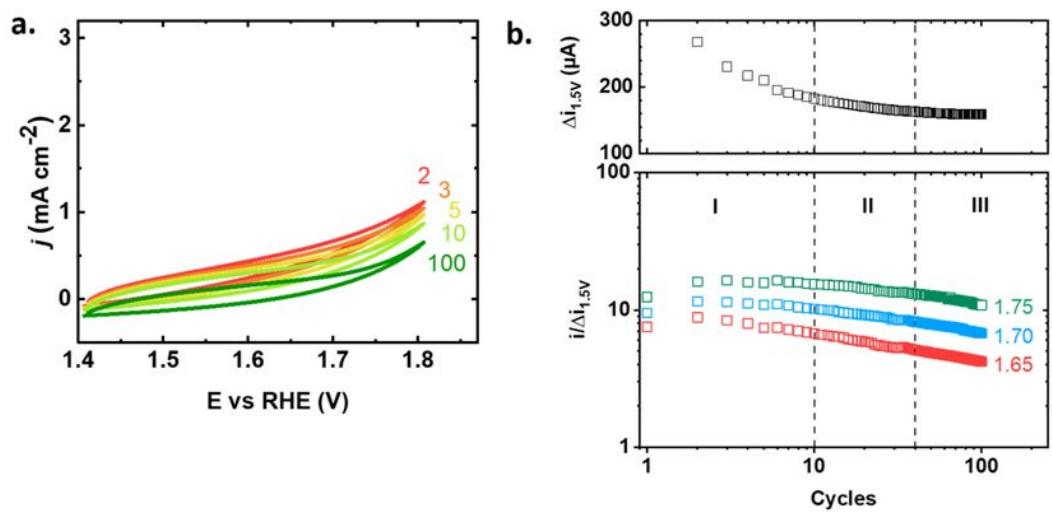

**Figure S14. a**. Series of CV of MnO$_x$ performed at the samples on graphite foil. CV with 100 mV s$^{-1}$ in 0.1 M NaOH. **b.** Average current ratios i/Δi$_{1.5V}$ as function of cycles at selected potentials of MnO$_x$. The data was evaluated during the first 100 cycles (open squares). The dashed lines delimit three different regions, labeled as: I, II and III.

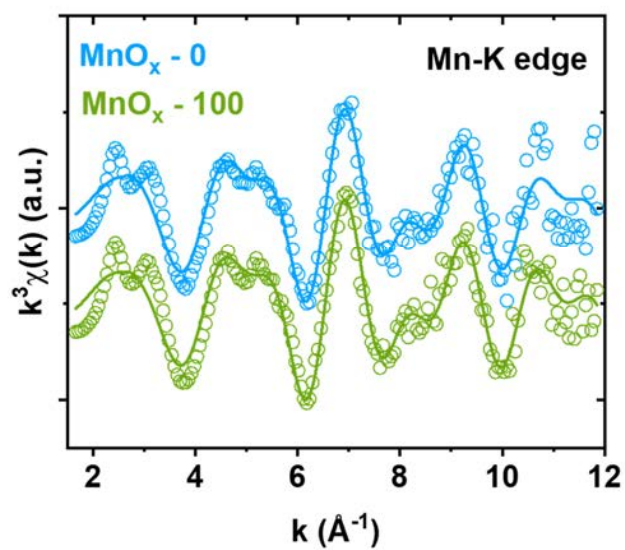

**Figure S15.** $K^3$-weighted EXAFS spectra of pristine $MnO_x$ ($MnO_x$ - 0) and after 100 cycles ($MnO_x$ - 100) recorded at the Mn-K edge. The solid lines represent the respective EXAFS simulations. The symbols represent the measurement.

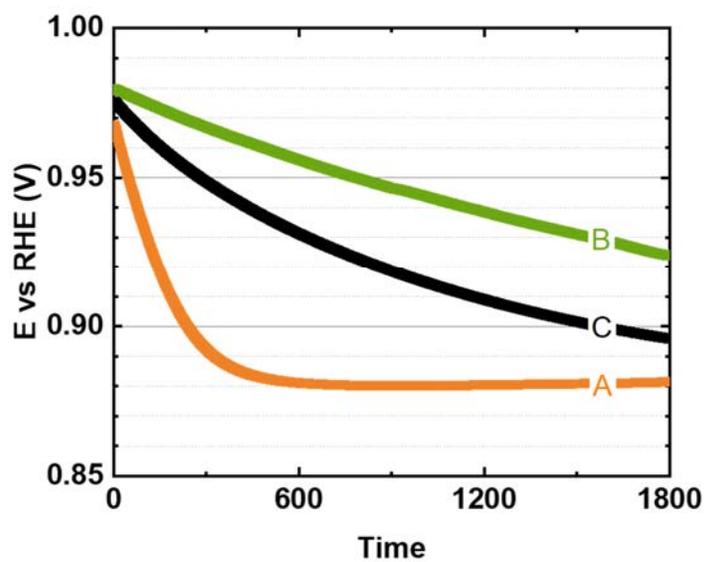

**Figure S16.** Open-circuit voltage profile during 30 minutes after 100 cycles of catalysis for three samples of $MnO_x$.